# Defect structures in MgB$_2$ wires introduced by hot isostatic pressing

X Z Liao[1], A Serquis[1], Y T Zhu[1], L Civale[1], D L Hammon[1], D E Peterson[1],

F M Mueller[1], V F Nesterenko[2] and Y Gu[2]

[1]Superconductivity Technology Center, MS K763, Los Alamos National Laboratory,

Los Alamos, NM 87545

[2]Department of Mechanical and Aerospace Engineering, University of California,

San Diego, La Jolla, CA 92093

Corresponding author:

Dr. Xiaozhou Liao

MST-STC, MS K763

Los Alamos National Laboratory

Los Alamos, NM 87545

USA

Phone:  505 665 6875

Fax:       505 667 2264

Email: xzliao@lanl.gov

**Abstract:**

The microstructures of $MgB_2$ wires prepared by the powder-in-tube technique and subsequent hot isostatic pressing were investigated using transmission electron microscopy. Large amount of crystalline defects including small angle twisting, tilting, and bending boundaries, in which high densities of dislocations reside, were found forming sub-grains within $MgB_2$ grains. It is believed that these defects resulted from particle deformation during the hot isostatic pressing process and are effective flux pinning centers that contribute to the high critical current densities of the wires at high temperatures and at high fields.

## I. Introduction

Magnesium diboride ($MgB_2$), with a superconducting critical temperature at 39 K [1], has caught the attention of scientists for its possibility of applications in magnets and electronic devices. Because of its low cost, low anisotropy, larger coherence lengths, and strong coupling across grain boundaries [2,3], $MgB_2$ may be used to replace low critical temperature ($T_c$) superconductors or Bi-based high temperature superconductors (BSCCO) in some applications in the temperature range of 4-30 K and magnetic fields of 0-12 T. For practical applications using $MgB_2$, it is essential to have high critical current density $J_c$ in high magnetic fields and this can be achieved through significantly increasing the flux pinning centers in $MgB_2$.

Many kinds of crystal defects, such as dislocation networks [4], stacking faults [5], grain boundaries [6,7,8], and non-superconducting second phases [9,10,11], can act as effective flux pinning centers in superconductors. A variety of techniques have been used to create crystal defects that act as local flux pinning centers in high-$T_c$ superconductors. They include chemical doping [12,13], and ion [14,15], neutron [16], pulse-laser [17], and electron [18,19] irradiation. Some of these techniques have also been successfully applied to the $MgB_2$ system [20,21].

In a previous paper [22], we presented the highest $J_c$ results ever reported on $MgB_2$ wires. The wires were prepared by the power-in-tube (PIT) technique and hot isostatic pressing (HIPing) [23]. We concluded that structural defects introduced by the HIPing are responsible for the very high $J_c$ in our HIPed wires and that HIPing is a very effective way of significantly increasing the flux pinning centers in $MgB_2$. In this paper, we present detailed transmission electron microscopy (TEM) observations of the defect structures in the $MgB_2$ wires introduced by the HIPing process.

## II. Experimental procedures

Commercial $MgB_2$ powder (Alfa Aesar, 98% purity) was first ball-milled for two hours before it was packed into stainless steel tubes. An as-drawn $MgB_2$ wire with the inner and outer diameters of 0.5 mm and 0.8 mm, respectively, was prepared by the powder-in-tube technique [24,25] and then cut into 4 inches long pieces, sealed at both ends by electric arc welding. The wire was HIPed at 900°C under an isostatic pressure of 200 MPa for 30 minutes. After the HIPing, the pressure was first reduced to ambient, and then the wires were cooled at 5°C/min to room temperature.

Samples for TEM investigation were prepared by mechanically grinding the $MgB_2$ wires to a thickness of about 30 μm and then further thinning to a thickness of electron transparency using a Gatan precision ion polishing system with $Ar^+$ accelerating voltage of 3.5 kV. TEM bright-field imaging and electron diffraction investigations were carried out using a Philips CM30 TEM working at 300 kV. High-resolution transmission electron microscopy (HREM) investigations were carried out using a JEOL 3000F TEM working at 300 kV.

### III. Results and discussions

TEM investigation of large areas of the HIPed wires and comparison with the structures of un-HIPed wires suggest that less porosity exists in the HIPed wires than in the un-HIPed wires. There are generally two kinds of areas in the HIPed wires: dense areas and areas with porosity. A typical example is shown in Fig. 1(a) in which a dense area is marked with A while an area with porosity is marked with B. Comparison at higher magnifications shows that the $MgB_2$ grain sizes in a dense area (Fig. 1 (b)) are normally smaller than those in an area with porosity (Fig. 1 (c)). This observation suggests that using $MgB_2$ powders with finer grain sizes is an effective way to further increase the density of a wire prepared by the powder-in-tube technique and therefore to get better connectivity. The observation also indicates that diffusion based mechanisms are important for densification of $MgB_2$ at least at the final stages of the densification. This density variation in the HIPed wires agrees with the observed variation of elastic properties in HIPed $MgB_2$ samples [26].

Each $MgB_2$ crystallite in the HIPed wires is normally divided into a few sub-grains with the sub-grain boundaries roughly parallel to (001) plane. Sub-grain boundaries can be easily identified by TEM when crystallites have their (001) paralleled to the electron beam direction. A typical example is shown in Fig. 2 in which sub-grain boundaries are presented clearly in three crystallites labeled with A, B and C, respectively. In order to understand the structural nature of the sub-grains in $MgB_2$ crystallites, HREM investigations were carried out on the sub-grain boundaries. Typical images are shown in Fig. 3. In Fig. 3 (a), the (001) (or [001]) of three sub-grains, marked I, II and III, respectively, are parallel to each other. The sub-grain boundaries among these three sub-grains are parallel approximately to (001) with some kinks seen at the boundaries (one of the kinks is indicated with a black arrow). It is seen from the magnified inserts under the labels I, II and III, respectively, in Fig. 3 (a) that (i) sub-grain I presents clear two-dimensional lattice fringes, implying a <100> zone-axis orientation; (ii) sub-grain III also shows two-dimensional fringes but with poor quality, indicating a minor angle off the <100> zone-axis, and (iii) sub-grain II displays only one-dimensional lattice fringes, suggesting a larger angle off the <100> zone-axis but the (001) of sub-grain II is still parallel to the (001) of sub-grains I and III. A specimen tilting experiment suggests that there is about 3º orientation difference between the <100> of sub-grains I and II. Combining the above observations, a conclusion is drawn that the sub-grains in Fig. 3 (a) were formed by small angle twisting around [001] with (001) as the twisting plane among regions I, II and III.

Tilting also plays an important role in the formation of the sub-grains. Figure 3 (b) shows an example of a twisting/tilting sub-grain boundary. Two black lines are drawn in Fig. 3 (b), with each line parallel to the (001) of sub-grains I and II, respectively, indicating a 5º tilt between the two neighboring sub-grains. The fact that sub-grain I in Fig. 3 (b) is on a <100> zone-axis (showing two-dimensional lattice fringes) while sub-grain II is off zone-axis (showing only one-dimensional lattice fringes) suggests some twisting between the two neighboring sub-grains. Therefore the boundary in Fig. 3 (b) is a combined twisting/tilting sub-grain boundary. The black arrows in Fig. 3 (b) are discussed later.

Because the axis of relative rotation among twisting sub-grains is parallel to [001] and the axis of relative rotation among tilting sub-grains lies in the plane of the tilting boundaries (i.e. (001)), the twisting nature of sub-grain boundaries can also be recognized from [001] selected area electron diffraction (SAED) patterns while the tilting nature can also be detected from <100> SAED patterns (note that a <100> is parallel to (001)). A typical example of a <100> SAED pattern taken from one $MgB_2$ crystallite is shown in Fig. 4. The diffraction spots in the pattern are all elongated, forming arcs. This pattern suggests tilting occurs and the accumulated tilting angle is up to about 21° in this specific crystallite, as measured from the two ends of a diffraction arc.

In addition to small angle twisting/tilting sub-grain boundaries, small angle bending sub-grain boundaries are also frequently seen in $MgB_2$ crystallites. A typical example is shown in a <100> zone-axis image of a $MgB_2$ crystallite in Fig. 5 in which two bending sub-grain boundaries, which show an abrupt image contrast change, are indicated by two large black arrows. Note that the bending can also be regarded as tilting relative to a plane other than the (001). Twisting/tilting sub-grain boundaries, which are similar to those shown in Fig. 3 and appear as parallel dark lines along (001), as indicated by small white arrows, are also seen in Fig. 5.

High densities of dislocations exist at small angle twisting/tilting sub-grain boundaries. For example, Fig. 3(b) shows dislocations in a twisting/tilting sub-grain boundary in which black-white contrast appears roughly periodically at an interval of about 2.5 nm, as marked with black arrows, along the boundary. At the center of each black-white contrast point is a dislocation.

According to the theory of dislocations at grain boundaries [27], each small-angle twisting/tilting grain boundary is made up of sets of uniformly spaced parallel and macroscopically straight dislocations. For the case of a small-angle (001) tilting grain boundary in $MgB_2$, there are two sets of dislocations at the boundary with spacings of $D_1 = b_1/(\theta\sin\phi)$ and $D_2 = b_2/(\theta\cos\phi)$, respectively, here $b_1$ and $b_2$ corresponds to the edge components of the two sets of dislocations, $\theta$ is the tilting angle and $\phi$ the angle between the boundary plane and the mean [001] direction of the two sub-grains. Therefore, the dislocation densities can be very high depending on the tilting angle $\theta$. For example, for a symmetrical ($\phi = 0º$) (001) tilting boundary with the tilting angle $\theta = 10º$ in $MgB_2$, the dislocation spacing would be only about 2 nm. For a small-angle (001) twisting grain boundary in $MgB_2$, two sets of dislocations form a crossed grid at the boundary with the spacing between dislocations of $D = b/\theta$ where b is the screw component of the dislocations and $\theta$ the twisting angle, for each set. Note that the above formulae are only valid for the case of equilibrium small-angle sub-grain boundaries. The dislocation density at a non-equilibrium grain boundary is higher than at an equilibrium boundary [28].

We have reported that the $J_c$ values of the HIPed wires at high fields are the highest in wires so far and are around one order of magnitude higher than those of the un-HIPed annealed wires [22]. This is obviously due to the improvement of flux pinning and reduction of porosity. At high fields, where the pinning enhancement due to the HIPing is more significant, the distance between vortices is small (~15nm at 8T, ~11nm at 16T) [22], so the defects involved must have a large density. Therefore, the short separation between dislocations seen in Fig. 3 (b) makes them good candidates. Dislocations are known to act as pinning centers both in conventional superconductors [29] and in epitaxial $YBa_2Cu_3O_7$ thin films and coated conductors [30,31], where they also form at low angle grain boundaries. In addition to the high density of defects in the HIPed wires, improvement in the connectivity in the HIPed wires is also believed to have contributed to the enhancement of $J_c$ in the HIPed wires.

Because the abovementioned defects are only seen in HIPed wires, it is believed that the high density of defects in HIPed wires was induced by the high strain and high temperature deformation of $MgB_2$ particles, which also results in pore collapse, during HIPing. These defect structures as well as the high density of dislocations suggest that $MgB_2$ grains were plastically deformed at high temperature by dislocation slip. As stated in the experimental section, the $MgB_2$ powder was ball milled at room temperature. However, $MgB_2$ is a brittle material with fracture toughness of 2.2 $MPa•m^{1/2}$ [26], which is comparable to alumina. As a result, no dislocation activity is expected during the room temperature ball milling. Therefore, the defects structures shown in Figs. 2, 3 and 5 can only be produced by particle deformation during the HIPing. During the HIPing at 900°, the $MgB_2$ particles need to be viscoplastically deformed so that they can accommodate each other to form porosity free aggregates. Dislocation slip appears to have played a critical role in the high temperature viscoplastic deformation, which generated the dislocation-related defect structure observed in this study. The high HIPing temperature activated dislocation slip. Other deformation mechanisms such as diffusion should also have played a role in the densification of $MgB_2$, as indicated by the dependence of local density on particle size.

**IV. Summary**

We have investigated the microstructures of a HIPed $MgB_2$ wire, which have demonstrated the highest $J_c$ value at high fields for $MgB_2$ wires, using TEM. Results suggest that high density of small angle twisting, tilting and bending sub-grain boundaries, at which high densities of dislocations reside, are responsible for the greatly improved flux pinning in the HIPed wires. It is believed that the defects are introduced by the high temperature particle deformation during the HIPing process. Our investigations also suggest that small grain sizes are beneficial to good connectivity or reducing porosity, indicating the importance of diffusion based mechanisms of densification.

**Acknowledgement**

**Figure captions**

Figure 1 (a) A low magnification TEM image showing two kinds of areas: a dense area marked with "A" and an area with porosity labeled with "B"; (b) a large magnification TEM image of the dense area and (c) a large magnification TEM image of the area with porosity.

Figure 2 A TEM image in which three crystallites, marked with A, B, and C, respectively, clearly show sub-grain boundaries roughly parallel to (001).

Figure 3 (a) An HREM image showing three sub-grains, marked with I, II and III, respectively, have their (001) (or [001]) parallel to each other. The HREM images of the three sub-grains are magnified and shown as inserts under labels I, II and III, respectively. A kink at the boundary of sub-grains II and III is marked with an arrow; (b) a high-resolution TEM image of a twisting/tilting sub-grain boundary. Black arrows indicate the centers of black-white contrast points at which dislocations are located. Two black lines, each parallel to the (001) of sub-grains I and II, respectively, are drawn together, showing a 5º small angle tilt between the two neighboring sub-grains.

Figure 4 A <100> electron diffraction pattern taken from a $MgB_2$ crystallite. Diffraction spots are elongated, form diffraction "arcs", because of tilting among sub-grains in the crystallite. The accumulated tilting angle is about 21°, as measured from the two lines drawn from the center of the diffraction pattern to the two ends of a diffraction arc.

Figure 5 A TEM image showing two small angle bending sub-grain boundaries, as indicated with two black arrows, respectively, in a $MgB_2$ crystallite. Twisting/tilting sub-grain boundaries seen in this crystallite are also pointed out by white arrows.

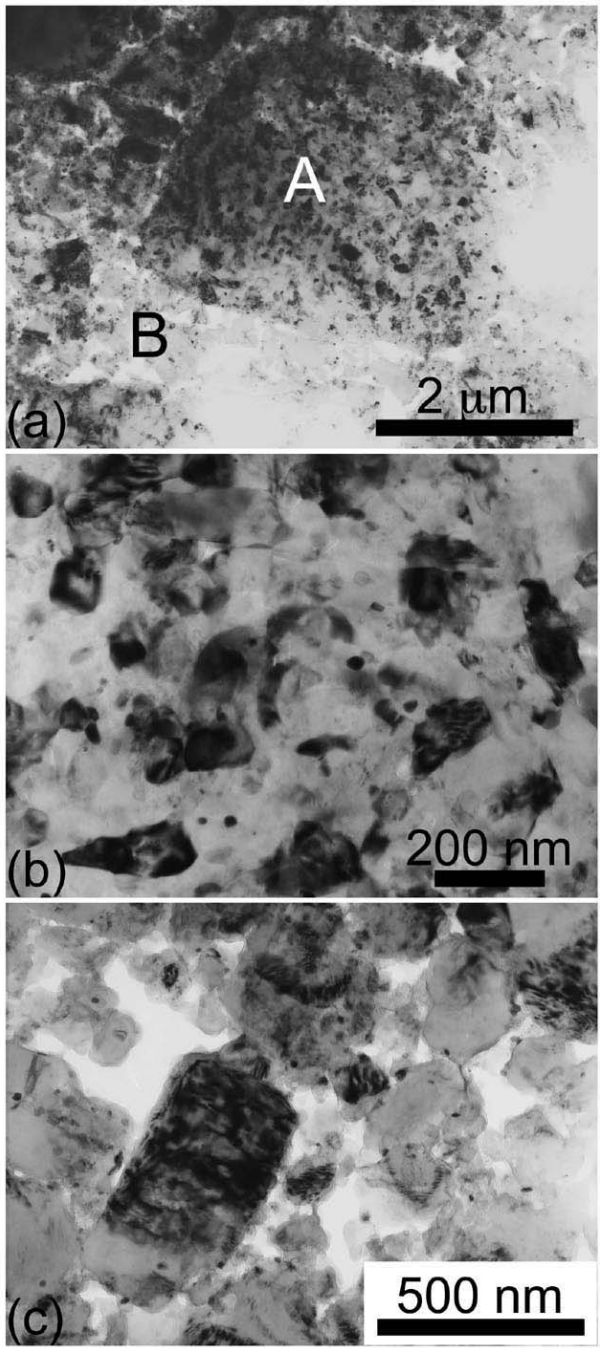

Fig. 2  (Liao et al)

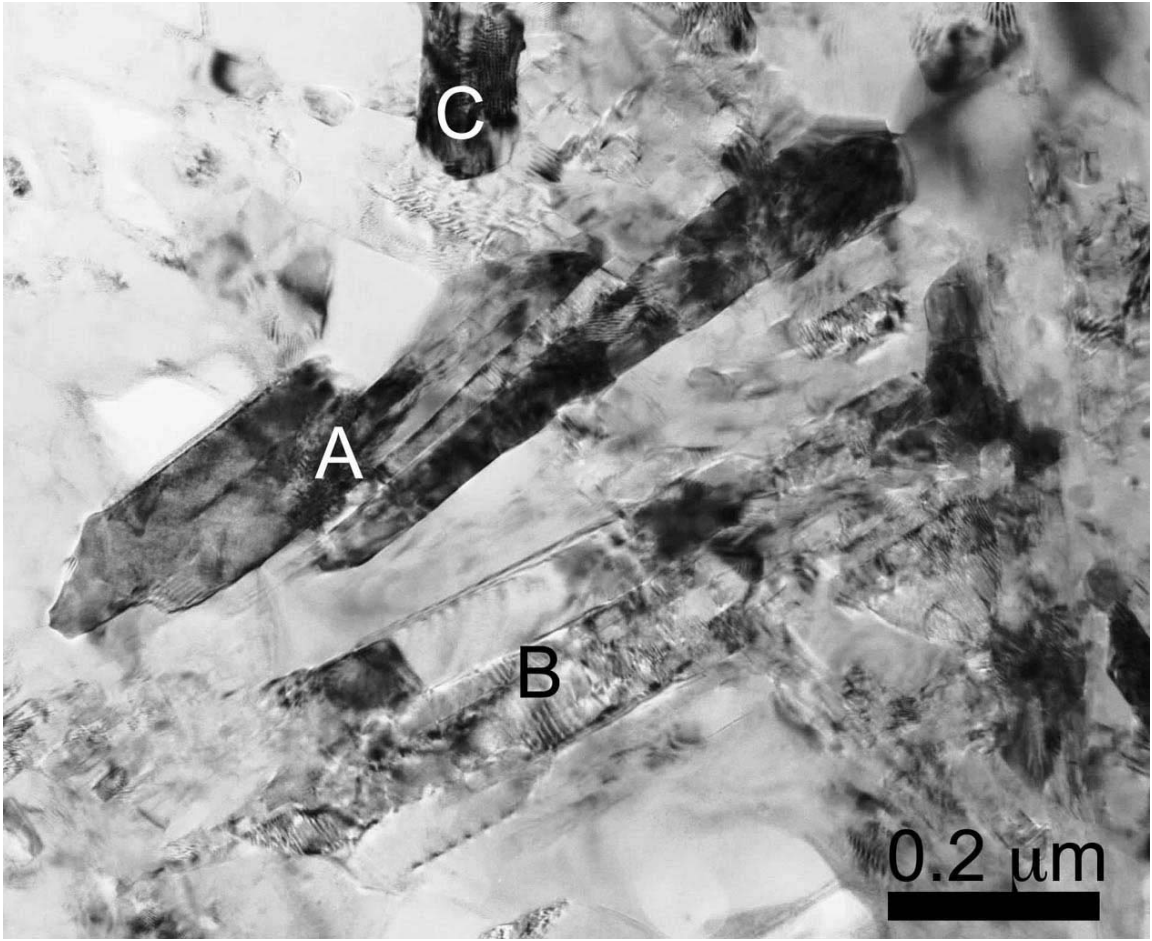

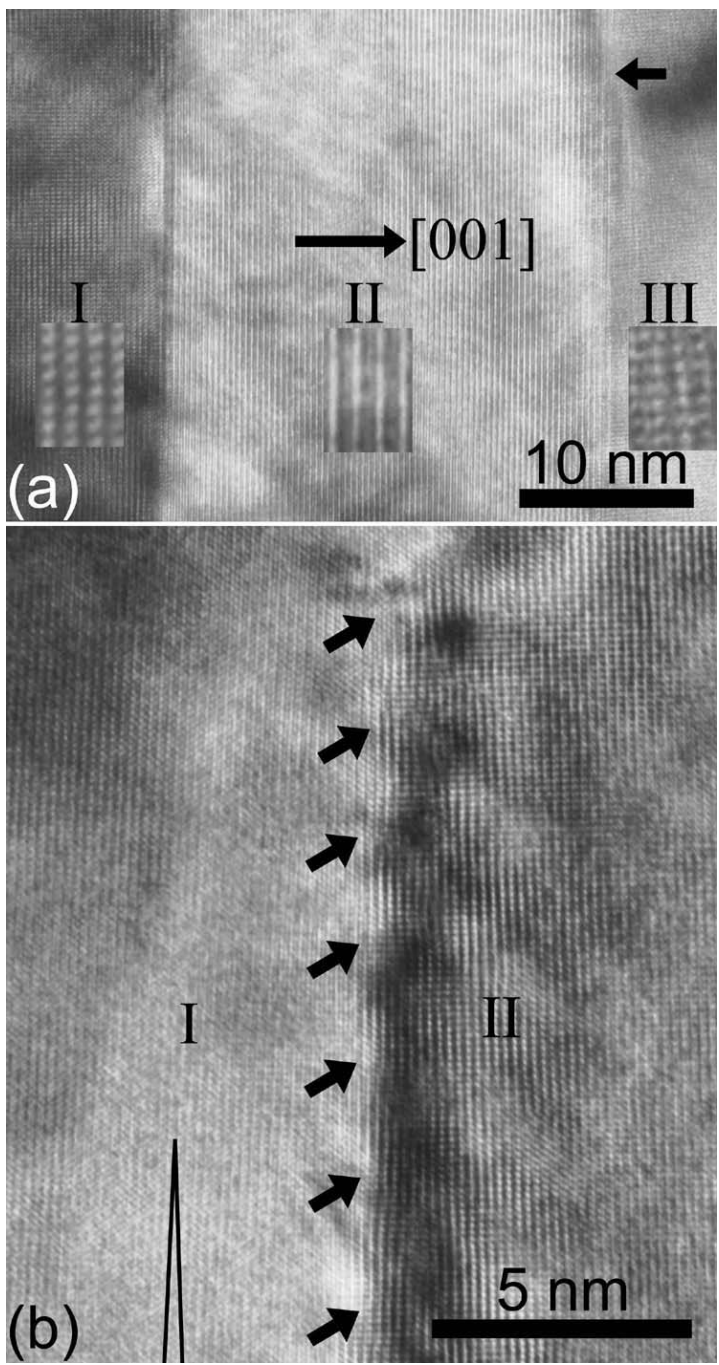

Fig. 4 (Liao et al)

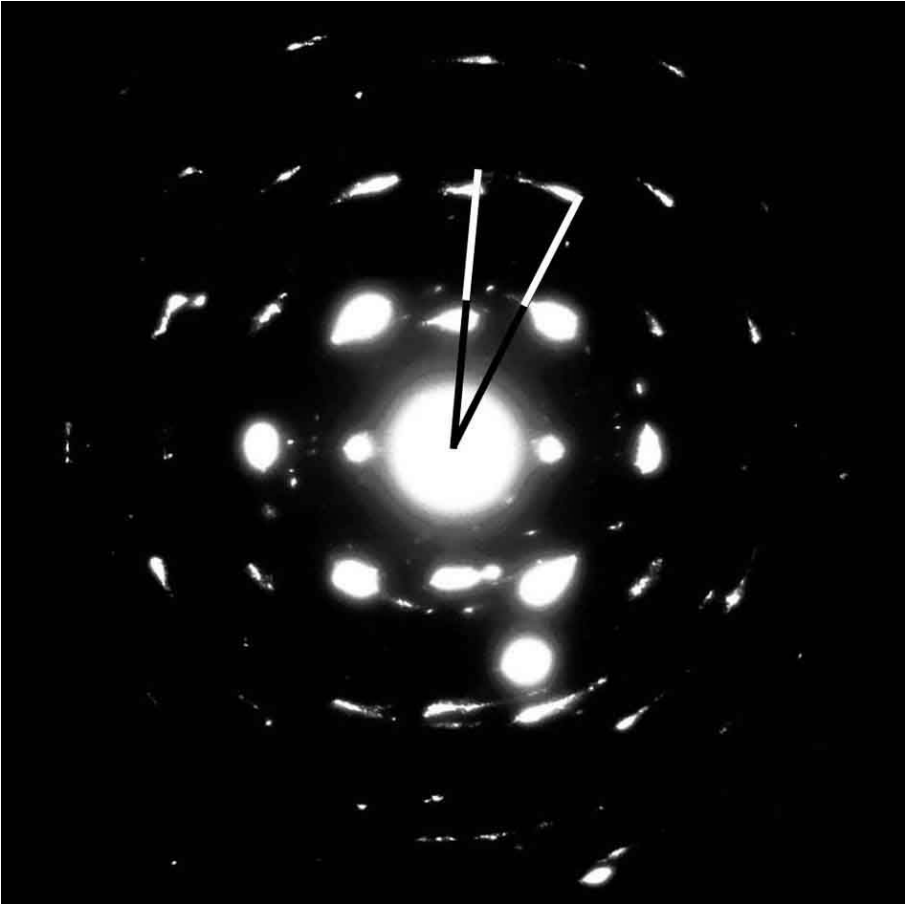

Fig. 5 (Liao et al)

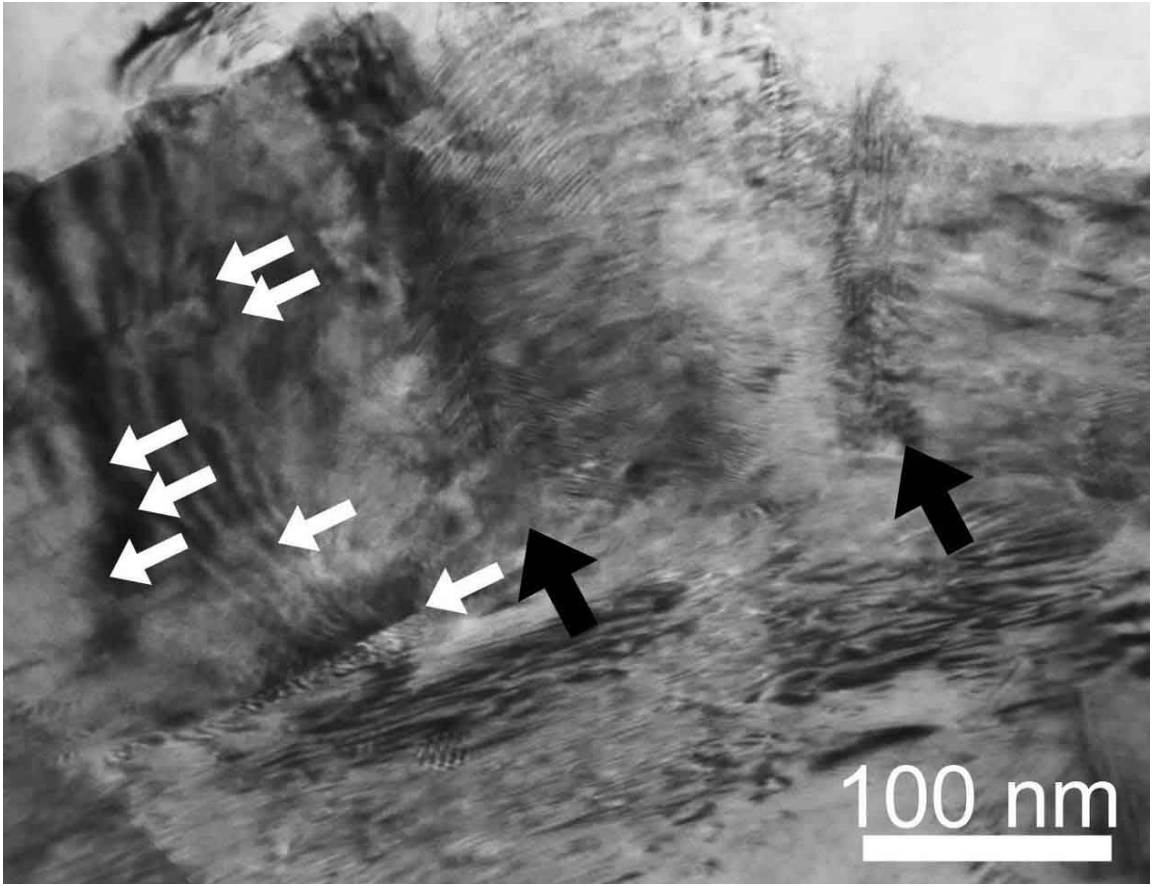